\documentclass[a4paper]{article}
\usepackage{algorithm, algpseudocode}
\usepackage{amsmath, amssymb}
\usepackage{array}
\newcolumntype{P}[1]{>{\centering\arraybackslash}m{#1}}
\usepackage{authblk}
\usepackage{booktabs}
\usepackage{braket}
\usepackage[labelfont=bf, font=footnotesize]{caption}
\usepackage{enumitem}
\usepackage{graphicx}
\usepackage[margin=1.2in]{geometry}
\usepackage{hyperref}
\usepackage{ragged2e} 
\usepackage[counterclockwise]{rotating}
\usepackage{url}
\usepackage[table]{xcolor}

\usepackage{cleveref} 

\hypersetup{
    colorlinks=true,
    linkcolor=blue,
    filecolor=magenta,      
    urlcolor=cyan,
    citecolor=blue
}

\title{\textbf{Supervised Similarity for Firm Linkages}}

\author[1, *]{Ryan Samson}
\author[2]{\authorcr Adrian Banner}
\author[1, 3]{Luca Candelori}
\author[4]{Sebastien Cottrell}
\author[5, 6, 7]{Tiziana Di Matteo}
\author[2]{Paul Duchnowski}
\author[1]{Vahagn Kirakosyan}
\author[2]{Jose Marques}
\author[1, 8]{Kharen Musaelian}
\author[9]{Stefano Pasquali}
\author[2]{Ryan Stever}
\author[1, 5, 8]{Dario Villani}
\affil[1]{Qognitive, Inc., Miami Beach, FL}
\affil[2]{Intech Investment Management LLC, West Palm Beach, FL}
\affil[3]{Wayne State University, Department of Mathematics, Detroit, MI}
\affil[4]{Deutsche Bank, New York, NY}
\affil[5]{King’s College London, Department of Mathematics, London, UK}
\affil[6]{Complexity Science Hub Vienna, Vienna, Austria}
\affil[7]{Centro Ricerche Enrico Fermi, Rome, Italy}
\affil[8]{Duality Group, Miami Beach, FL}
\affil[9]{Domyn, New York, NY}
\affil[*]{Primary Investigator (\href{mailto:ryan.samson@qognitive.io}{ryan.samson@qognitive.io}), remaining authors presented in alphabetical order.  The authors also wish to thank Hongli Lan for his contributions to this article.
\authorcr The views expressed here are those of the authors alone and not of Deutsche Bank, Domyn, Duality Group, Intech Investment Management, or Qognitive.  
\authorcr The contents of this article do not constitute investment advice.}
\date{\today}

\begin{document}
\maketitle

\begin{abstract}
We introduce a novel proxy for firm linkages, Characteristic Vector Linkages (CVLs).  We use this concept to estimate firm linkages, first through Euclidean similarity, and then by applying Quantum Cognition Machine Learning (QCML) to similarity learning.  We demonstrate that both methods can be used to construct profitable momentum spillover trading strategies, but QCML similarity outperforms the simpler Euclidean similarity.
\end{abstract}

\section{Introduction}

Prior literature has explored the use of fundamental information as a proxy for firm linkages.  If  investors have limited attention, then news impacting the price of a firm may only slowly be incorporated into prices of related firms, leading to return predictability across firms.  Indeed, for many such firm linkages it has been shown that lagged returns of a firm are predictive of future returns for firms which are more similar to it.  This effect is sometimes referred to as a momentum spillover effect, or a lead-lag strategy.  Momentum spillover has been documented for similarities formed from a variety of fundamental information including industry \cite{Moskowitz_Grinblatt_1999}, supply chain \cite{Cohen_Frazzini_2008}, analyst coverage \cite{Ali_Hirshleifer_2019}, and geography \cite{Pirinsky_2006}, among others.

Unrelated literature explores the application of machine learning techniques to the learning of similarity relations between securities, often with the goal of clustering securities for risk management, signal generation, or portfolio construction.  See e.g. the literature review in \cite{Hayette_Salah_2025} for examples of classification and clustering techniques, \cite{Vamvourellis_2023} for a demonstration of how embeddings from Large Language Models can be used to extract company similarity relations, or \cite{Bartram_et_al_2021} for a more general review of machine learning applications in finance.

More recent work has begun to explore the use of supervised learning techniques to extract similarity relationships.  \cite{Jeyapaulraj_et_al_2022} proposed a supervised similarity framework based on extracting distance metrics from Random Forest, and demonstrated an application to clustering corporate bonds.  \cite{Rosaler_et_al_2025} extended work on the corporate bond clustering problem, demonstrating the application of Quantum Cognition Machine Learning \cite{Musaelian_et_al_2024, Samson_et_al_2024, Candelori_et_al_2025}, a novel paradigm for both supervised and unsupervised learning tasks rooted in the mathematical formalism of quantum theory, to distance metric learning.

Here, we introduce a novel proxy for firm linkages, which we name Characteristic Vector Linkages.  Distance based on these linkages can most simply be derived from the Euclidean distance of a vector of firm characteristics, which can then be used to form a profitable momentum spillover trading strategy.  Next, we demonstrate how an approach similar to that discussed in \cite{Rosaler_et_al_2025} can be used to learn distance relationships across equity securities, which can in turn be used to enhance the momentum spillover trading strategy derived from the simpler Euclidean distance.

\section{Characteristic Vector Linkages and Euclidean Distance}
 \label{section:cvl}

 Academic and practitioner literature has identified a large number of stock characteristics which are widely used as cross-sectional predictors of risk or returns\footnote{See e.g. \cite{Hou_2020} or \cite{Jensen_2023}, both of which evaluate the robustness of various features proposed in the financial literature, and also serve as excellent surveys of such features / characteristics}.  The values of these characteristics, or factors, give us small but important pieces of information about the underlying firms.  We propose that, across a broad set of characteristics, the more similar the factor scores of any pair of firms, the more economically similar the underlying pair of firms truly is, and the more likely they are to display momentum spillover effects.  In other words, similarity of sets of factor scores can be used to measure the strength of firm linkages.  We name such purported factor / characteristic derived linkages Characteristic Vector Linkages (CVLs).  

 Define a characteristic observed at time $t$ as $x_{t,j}^c$, where $c = 1,2,\dots,C$ is the index of a characteristic and $j = 1,2,\dots,J$ is the index of a firm.  $\mathbf{x}_{t}^c \in \mathbb{R}^J$ is the vector over all firms for a single characteristic, $\mathbf{x}_{t,j} \in \mathbb{R}^C$ is the vector over all characteristics for a single firm, and $\mathbf{x}_t \in \mathbb{R}^{J\times C}$ is the matrix over firms and characteristics (we use similar index-omission notation throughout this article).  Distance of factor scores could be defined in a variety of ways, but can simply and naturally be expressed as the Euclidean distance between two firm characteristic vectors.  The Euclidean distance between firms $i$ and $j$ at time $t$ is: 
\begin{equation}
    \label{eq:euclidean_distance}
    D_{Euclidean}(\mathbf{x}_{t,i}, \mathbf{x}_{t,j}) = \sqrt{(\mathbf{x}_{t,i} - \mathbf{x}_{t,j})^\intercal(\mathbf{x}_{t,i} - \mathbf{x}_{t,j})}
\end{equation}
which we can then transform to a similarity measure on $[0, 1]$ as will be explained in section \ref{section:distance_to_similarity}.  

Euclidean distance is a natural choice in this context, as the magnitude of the factor scores and their differences carry meaning.  By contrast, consider a normalized distance metric such as cosine distance:
\begin{equation}
    \label{eq:cosine_similarity}
    D_{cosine}(\mathbf{x}_{t,i}, \mathbf{x}_{t,j}) = 1- \frac{\mathbf{x}_{t,i} \cdot \mathbf{x}_{t,j}} {||\mathbf{x}_{t,i}||\; ||\mathbf{x}_{t,j}||}
\end{equation}
where input vectors are scaled by their norms.  To understand why this is problematic, consider the extreme case of firm $i$ with $\mathbf{x}_{t,i}^c = \epsilon\; \forall\; c$ and firm $j$ with $\mathbf{x}_{t,j}^c = -\epsilon\; \forall\; c$.  These firms have nearly identical factor scores and thus our distance metric should return a small value.  In this case, equation \ref{eq:euclidean_distance} gives $D_{Euclidean}(\mathbf{x}_{t,i}, \mathbf{x}_{t,j}) = 2\epsilon\sqrt{C}$, a small value, while equation \ref{eq:cosine_similarity} gives $D_{cosine}(\mathbf{x}_{t,i}, \mathbf{x}_{t,j}) = 2$, the maximum distance possible.

\section{Quantum Cognition Machine Learning Distance}
 \label{section:qcml}

\subsection{Quantum Cognition Machine Learning Review}
\label{section:qcml_review}

Quantum Cognition Machine Learning (QCML) \cite{Musaelian_et_al_2024, Samson_et_al_2024, Candelori_et_al_2025} has recently been proposed as a new form of machine learning based on quantum cognition (see \cite{Porthos_Busemayer_2022} for a recent survey of quantum cognition).  QCML models learn a representation of the data into quantum states. Recall that in quantum mechanics a {\em state} is a vector of unit norm in a Hilbert space, and is represented in bra-ket notation by a ket $\ket{\psi}$. The inner product of two states $\ket{\psi_1}$, $\ket{\psi_2}$ is represented by a bra-ket $\braket{\psi_1|\psi_2}$. The expectation value of a Hermitian operator $O$ (i.e. a quantum observable) on a state $\ket{\psi}$ is denoted by $\bra{\psi}O\ket{\psi}$, representing the expected outcome of the measurement corresponding to $O$ on the state $\ket{\psi}$.

For QCML, we use $\ket{\psi} \in \mathbb{C}^N$, operators $O \in \mathbb{C}^{N\times N}$ and the standard Hermitian inner product, so kets correspond to column vectors, and bras to their conjugate transpose.  Combinations of bras, kets, and operators are then interpreted using standard matrix multiplication rules.  Hereafter, we omit the ket from $\psi$ unless part of such an expression.
 
QCML starts by defining an error Hamiltonian over data and observables.  We have flexibility in how we define the Hamiltonian, so long as the result is non-negative and Hermitian.  In this article we have defined the error Hamiltonian as:
\begin{equation}
    \label{eq:square_hamiltonian}
    H(\mathbf{x}_{t,j}, \{A_c\}) = \sum_c (A_c - x_{t,j}^c \cdot I)^2
\end{equation}
where $\mathbf{x}_{t,j}$ is as defined in section \ref{section:cvl}, $\{A_c\} \in \mathbb{C}^{N \times N} $ is a set of $C$ Hermitian observable operators which must be learned, and $I$ denotes the $N \times N$ identity matrix.  

Each of the $C$ quantum observables $\{A_c\}$ can be viewed as a ``quantization'' of a corresponding feature of the original $C$-dimensional data set. The vector $\mathbf{x}_{t,j}$ is then mapped to the ground state $\psi_{t,j}$ of the error Hamiltonian,\footnote{The ground state $\psi_{t,j}$ is the eigenstate associated to the lowest eigenvalue of the error Hamiltonian, and is also called the quasi-coherent state} giving a representation of data into quantum states (i.e., unit norm vectors in a complex Hilbert space). Conversely, given an arbitrary $N$-dimensional quantum state $\psi_{t,j}$, we can define its ``position'' to be the $C$-dimensional vector:
\begin{equation}
\widehat{\mathbf{x}}_{t,j} = (\bra{\psi_{t,j}}{A_c}\ket{\psi_{t,j}})_{c} \in \mathbb{R}^C, 
\end{equation}
which in quantum theory represents the expected outcome of measuring the quantum observables $A_c$ on $\psi_{t,j}$. In this way, given a set of quantum observables $\{A_c\}$, we have a way to map data into quantum states by sending $\mathbf{x}_{t,j}$ to its ground state $\psi_{t,j}$, and we can also retrieve information about a quantum state $\psi_{t,j}$ by taking its position $\widehat{\mathbf{x}}_{t,j}$. 

In an unsupervised setting, training a QCML model involves iterative updates to the estimated observables $\{\widehat{A}_c\}$ so that the ground states $\psi_{t,j}$ of the error Hamiltonian ``cohere'' to the data. That is, the distance between $\mathbf{x}_{t,j}$ and its position $\widehat{\mathbf{x}}_{t,j}$ is minimized, as well as the variance of the measurement.  The overall ground state energy of the error Hamiltonian (with $H$ defined in equation \ref{eq:square_hamiltonian}):

\begin{equation}
\label{eqn:energy}    
E_0(\mathbf{x}_{t,j}, \{\widehat{A}_c\}) = \bra{\psi_{t,j}}H\ket{\psi_{t,j}}
\end{equation}
is equivalent to the sum of this distance and the variance of the measurement.  See \cite{Candelori_et_al_2025} for a detailed discussion of this decomposition.  As discussed in \cite{Samson_et_al_2024}, ``targets'' can be included in the set of $C$ operators, and ground states and expectations can be taken over any subset of operators once the model is trained. 

In the supervised setting, which is the primary focus of this article, the training process is slightly different. The target variable $y_{t,j} \in \mathbb{R}$ is assigned an $N$-dimensional quantum ``forecast'' observable $B$, and given a data vector $\mathbf{x}_{t,j}$ the corresponding forecast is given by:
\begin{equation}
\label{eqn:yhat}    
\widehat{y}_{t,j} = \bra{\psi_{t,j}}{B}\ket{\psi_{t,j}}.
\end{equation}
During the training process, the estimated quantum observables $\{\widehat{A}_c\}$ and $\widehat{B}$ are updated at each iteration so as to minimize the mean squared error $\sum_{t,j} (\widehat{y}_{t,j} - y_{t,j})^2$.  This can easily be extended to the case of multiple target variables, and we can also add to the loss any of the components of $E_0(\mathbf{x}_{t,j}, \{\widehat{A}_c\})$ which we also wish to minimize.  For the results in this article we use:
\begin{equation}
\label{eqn:loss}    
Loss_{t,j} = f(y_{t,j}, \mathbf{x}_{t,j},\{\widehat{A}_c\}, \widehat{B},w) = (\widehat{y}_{t,j} - y_{t,j})^2 + w*\sum_c(\widehat{x}_{t,j}^c - x_{t,j}^c)^2
\end{equation}
where $w$ is a hyperparameter influencing how strongly the ground states cohere to the input data.  This form of the loss is not essential - for example, our empirical results in subsequent sections were similar, albeit slightly weaker, with $w=0$ in equation \ref{eqn:loss} or with the unsupervised approach (minimizing equation \ref{eqn:energy}, with target included in $\{A_c\}$).

The training algorithm can be summarized as follows:

\begin{algorithm}
    \caption*{\bf{QCML univariate regression model training with input bias penalty}}
    \begin{algorithmic}
    \State \textbullet\ Randomly initialize estimated feature operators $\{\widehat{A}_c\}$ and target operator $\widehat{B}$.
    \State \textbullet\ Iterate over training data until desired convergence:
    \begin{algorithmic} \item
        \begin{algorithmic}[1]
        \State Generate error Hamiltonian $H(\mathbf{x}_{t,j}, \{\widehat{A}_c\})$ (equation \ref{eq:square_hamiltonian})
        \State Holding $\{\widehat{A}_c\}$ constant, find the ground state $\psi_{t,j}$ of $H(\mathbf{x}_{t,j}, \{\widehat{A}_c\})$
        \State Calculate gradients w.r.t $\{\widehat{A}_c\}$ and $\widehat{B}$ of the loss function (equation \ref{eqn:loss}): $\sum_{t,j} Loss_{t,j}$
        \State Update $\{\widehat{A}_c\}$ and $\widehat{B}$ via gradient descent
        \end{algorithmic}
    \end{algorithmic}
    \end{algorithmic}
\end{algorithm}

The specifics of each of these steps will depend on the choice of parametrization for the operators $\{A_c\}$ and $B$. For results in this article, each operator is parameterized as a sum $O_1 + iO_2$, where $O_1$ (respectively, $O_2$) is a real symmetric (real antisymmetric) matrix. The dimension of the Hilbert space $N$ is a hyperparameter of the algorithm. Larger values of $N$ typically lower the loss, but could lead to overfitting and worse out-of-sample performance, while lower dimensions tend to have higher bias and lower variance \cite{Candelori_et_al_2025}. Results in this article were obtained using $N=12$, choices of $N \in [4, 32]$ have been seen to give optimal cross-validated accuracy for a wide variety of learning problems. 

\subsection{QCML Distance}
\label{section:qcml_distance}

As explained in section \ref{section:qcml_review}, QCML represents each observation as a quantum state. This allows one to have a natural notion of proximity between observations, since proximity between quantum states can be defined as {\em quantum fidelity} \cite[III.9]{Nielsen_et_al_2000}
\begin{equation}
\label{eqn:quantum_fidelity}
f(\psi_1, \psi_2) = \lvert \braket{\psi_1|\psi_2} \rvert ^ 2,
\end{equation}
which can be interpreted as the probability of identifying the state $\psi_1$ with the state $\psi_2$, when performing a quantum measurement designed to test whether a given quantum state is equal to $\psi_2$ (or vice versa).

Fidelity can be transformed to distance in a number of ways.  Here, we choose to use the Bures distance defined in \cite{Spehner_2017}:
\begin{equation}
\label{eqn:quantum_distance}
D_{QCML}(\mathbf{x}_{t,i}, \mathbf{x}_{t,j}) = \sqrt{2-2 \cdot \sqrt{f(\psi_{t,i}, \psi_{t,j})}}= \sqrt{2-2 \cdot \lvert \braket{\psi_{t,i}|\psi_{t,j}} \rvert}
\end{equation}
where vectors $\mathbf{x}_{t,i}, \mathbf{x}_{t,j}$ are mapped to ground
states $\psi_{t,i}, \psi_{t,j}$ as explained in section \ref{section:qcml_review}.  We chose the Bures distance for its similarity to Euclidean distance, noting that all ground states $\psi$ are unit norm.  Empirical results are not particularly sensitive to this choice of distance. For example, we obtained nearly identical results using the geodesic distance defined in \cite{Bengtsson_2006} as: $\arccos{\lvert \braket{\psi_{t,i}|\psi_{t,j}} \rvert}$.  We could also use fidelity directly as a similarity measure, rather than converting states to a distance and then converting distance to similarity.  However, directly using these probabilities results in similarities which are too diffuse.  Making these transformations allows us to form similarities with distributions comparable to those obtained using the Euclidean distance measure, allowing for a more fair comparison of the two.

\section{Converting Distance to Similarity}
 \label{section:distance_to_similarity}

Given a notion of distance, we form a similarity matrix following a similar approach to that outlined in \cite{Ng_2001} for affinity matrices:

\begin{equation}
\label{eq:similarity}
S(\mathbf{x}_{t,i}, \mathbf{x}_{t,j}) = 
\begin{cases}
e^{-\gamma \cdot D(\mathbf{x}_{t,i}, \mathbf{x}_{t,j})^2} & \text{if } i \ne j \\
0 & \text{otherwise}
\end{cases}
\end{equation}
Using this definition and our distance definitions in equations \ref{eq:euclidean_distance} and \ref{eqn:quantum_distance}, we can compute $J \times J$ similarity matrices by date, for all $i$, $j$ firm pairs.

Note that, in contrast to the Euclidean similarity between two vectors, the QCML similarity is a learned measure, since the representation of the data in quantum states $\psi_t$ has been optimized using the training targets.  Thus, in the context of QCML, the similarity measure is estimated on the data via the following steps:

\begin{enumerate}
    \item For $C$ input characteristics, learn a set of operators $\{A_c\}$ on a set of training data as described in section \ref{section:qcml_review}
    \item Compute $D_{QCML}(\mathbf{x}_{t,i}, \mathbf{x}_{t,j})$ as defined in equation \ref{eqn:quantum_distance}.  This requires first generating the error Hamiltonians $H(\mathbf{x}_{t,j})$ from equation \ref{eq:square_hamiltonian} and finding the corresponding ground states $\psi_{t,j}$ as described in section \ref{section:qcml_review}
    \item Compute the $J \times J$ similarity matrix $S_{QCML}(\mathbf{x}_t)$ with elements defined in equation \ref{eq:similarity}
\end{enumerate}

Our similarity matrix in equation \ref{eq:similarity} also requires a scaling parameter $\gamma$, which influences the concentration of the similarity values.  For the results which follow in this paper, we use $\gamma_{Euclidean}=1$ when computing Euclidean similarity, after verifying that results are not sensitive to this parameter and a value of $1$ gives close to optimal results.  When computing QCML similarity, we choose $\gamma_{QCML}=16$, which results in $\gamma_{Euclidean} \cdot D_{Euclidean}(\mathbf{x}_t)^2$ and $\gamma_{QCML} \cdot D_{QCML}(\mathbf{x}_t)^2$ having approximately matching median values over the training data.

\section{CVL Momentum Spillover Strategies}

\subsection{Overview}

We now illustrate how to apply our concept of CVLs to form a profitable momentum spillover trading strategy, using an example set of characteristics.  We first demonstrate the efficacy of the simple Euclidean similarity measure. We then show how QCML similarity can be used with the same input characteristics to learn similarity relationships across firms, which can in turn be used to enhance the momentum spillover trading strategy.

\subsection{Data and Features}
\label{section:data}

We collect daily data on a dynamic set of US firms screened for the largest size, liquidity and maturity at each point in time from October 2017 through June 2024.  The average number of names available per trading day in this universe is 1,500, but the set of names changes over time to reflect the most mature and liquid names available. This universe of securities is used for both model training and subsequent evaluation.  

We create a broad set of features to be used as inputs for our characteristic vectors.  We do not take a view on what characteristics are best suited for this task, but here we focus on accounting and valuation ratios, as such characteristics are widely used, easy to compute, and fairly stable.  Accounting values are averaged over the most recent 4 quarters, unless otherwise stated.  All input characteristics are demeaned by GICS Industry Code to avoid characteristics linking firms based on industry biases.  Characteristics are also cross-sectionally z-scored, and winsorized at the $1^{st}/\,99^{th}$ percentiles. We use market data sourced from Bloomberg and accounting data sourced from S\&P Capital IQ:

\begin{itemize}[align=left,leftmargin=10ex,labelsep=1ex]
\justifying

  \item[\textbf{Accounting Liquidity:}] (0.5 * Total Assets + 0.25 * Cash + 0.25 * Current Assets - 0.5 * Intangible Assets), scaled by Total Assets \cite{Ortiz_2014}

  \item[\textbf{Accruals:}] Sign flipped 4 quarter change of (Total Assets - Working Capital - Total Liabilities - Long Term Investments + Long Term Debt), scaled by Total Assets \cite{RichardsonBSAccruals, SloanAccruals}

  \item[\textbf{Book to Price:}] Common Equity, scaled by Market Cap \cite{Rosenberg_1985}
  
  \item[\textbf{Cash to Assets:}] Cash, scaled by Total Assets \cite{Palazzo_2012}
  
  \item[\textbf{Earnings to Price:}] Net Income, scaled by Market Cap \cite{Basu_1983}
  
  \item[\textbf{EBITDA to TEV:}] Earnings Before Interest, Taxes, Depreciation, and Amortization, scaled by Total Enterprise Value \cite{LoughranEbitdaTEV}

  \item[\textbf{Implied Equity Duration:}] Defined in \cite{Dechow_et_al_2004}, a measure which adapts the concept of bond duration to equities.  We use the forecasting parameters listed by the authors in Table 1, Panel B.
  
  \item[\textbf{Leverage:}] Total Debt, scaled by Total Assets \cite{bhandari_1988}
  
  \item[\textbf{Net Leverage:}] Total Debt - Cash and Short Term Investments, scaled by Total Assets \cite{penman_2007}
  
  \item[\textbf{Net Operating Assets:}] Long-Term Debt + Current Portion of Long-Term Debt + Short-term Borrowings + Minority Interest in Equity + Preferred Equity + Total Common Equity - Non-Stock Preferred Equity - Total Cash And Short Term Investments - Total Other Investments, scaled by Total Assets \cite{Hirshleifer_2004}
  
  \item[\textbf{Operating Cash-flows:}] Cash from Operations, scaled by Total Assets \cite{desai_2004}
  
  \item[\textbf{Operating Efficiency:}] Revenues, scaled by Total Assets \cite{SolimanDupont}
  
  \item[\textbf{Profit Margin:}] Net Income, scaled by Revenues \cite{SolimanDupont}

  \item[\textbf{Real Estate Ratio:}] Assets under Capital Lease + Assets under Operating Lease + Buildings, scaled by gross Property, Plant, and Equipment \cite{Tuzel_2010}
  
  \item[\textbf{Revenue to Price:}] Revenues, scaled by Market Cap \cite{BarbeeRevP}
  
  \item[\textbf{Tax Ratio:}] Earnings Before Taxes, including Unusual Items, scaled by Net Income to Company \cite{Lev_2004}

\end{itemize}

We also create the following set of controls for our portfolio construction process, using market data sourced from Bloomberg, and accounting and analyst ratings data sourced from S\&P Capital IQ.  All controls except GICS Dummies are winsorized at the $1^{st}/\,99^{th}$ percentiles:

\begin{itemize}[align=left,leftmargin=10ex,labelsep=1ex]
\justifying
  
  \item[\textbf{Analyst Connected Stock Momentum:}] Momentum spillover signal formed on the basis of shared stock analyst coverage, following the approach in \cite{Ali_Hirshleifer_2019}.  We create 5 versions of this control, each with input returns matched to the horizon of the input returns for the CVL signal being tested (see section \ref{section:signal_construction} for details of our CVL signal construction).  All versions are demeaned by GICS Industry Code and cross-sectionally z-scored
  
  \item[\textbf{Analyst Coverage:}] Number of analysts covering a stock, cross-sectionally z-scored \cite{parsons_2020}\footnote{As detailed in \cite{parsons_2020}, analyst coverage serves as an important proxy for investor attention, and some previously documented spillover signals are substantially weaker when controlling for analyst coverage}
  
  \item[\textbf{Beta:}] Rolling 252-day time series estimated Beta to S\&P 500 Index \cite{CAPM, FamaMacBeth, FF3}
  
  \item[\textbf{Momentum:}] Returns from 21 days ago to 252 days ago, cross-sectionally z-scored \cite{JTMom, CarhartMom}
   
  \item[\textbf{Short Term Reversal:}] Returns from 1 day ago to 100 days ago, exponentially weighted with a 10 day half life, returns within 1 day of earnings announcements replaced by beta-adjusted market returns, demeaned by GICS Industry Code, cross-sectionally z-scored  \cite{Rosenberg_1985}
  
  \item[\textbf{Size:}] Log of Market Cap, cross-sectionally z-scored \cite{BanzSize, FF3}

  \item[\textbf{Subindustry Momentum:}] Log of market cap average weighted returns for the firm's GICS subindustry from 1 day ago to 200 days ago, exponentially weighted with a 20-day half-life, cross-sectionally z-scored \cite{Moskowitz_Grinblatt_1999}
  
  \item[\textbf{Revenue to Price:}] As defined above
  
  \item[\textbf{GICS Dummies:}] Dummy variables formed based on GICS Industry Group \cite{AsnessIA, AsnessEtAlIA}

\end{itemize}

\subsection{Similarity Construction}
\label{section:setup}

As explained in section \ref{section:distance_to_similarity}, our Euclidean similarity measure is not a learned measure and requires no training.  Given the inputs defined in section \ref{section:data}, we can directly compute $S_{Euclidean}(\mathbf{x}_t)$ on each day the measure is needed during the test period (January 2014 through June 2024).  In contrast, our QCML similarity measure requires first training a QCML model, as explained in section \ref{section:distance_to_similarity}.  

QCML is trained using daily data from October 2007 through August 2013.  For model training, we are required to choose a target variable.  Note that the trained QCML model can be used to directly forecast the chosen target using equation \ref{eqn:yhat}.  However, in this article we are not interested in the efficacy of the direct forecast, and thus we do not include herein any analysis of its performance.  Instead, we will show that the ground states $\psi_{t,j}$ can be used to create a superior distance measure using the technique outlined in section \ref{section:qcml_distance}.  

It stands to reason that stocks which have similar future returns driven by observed characteristics are also likely to display stronger future co-movement due to similarity derived from those characteristics. Therefore, we choose as target variable 63-day forward returns, cross-sectionally z-scored.  Other choices of target could be made.  For example, future revenue / profitability growth or earnings surprises may more tightly link stocks for fundamental reasons.

Inputs are defined in section \ref{section:data}, and are the same inputs used for the Euclidean similarity measure.  To create more robust similarity measures, we train 50 unique QCML models with varying seeds.  Each seed randomly selects 10\% of available training names to use.  We form five distinct sub-groups of training dates (equally spaced), and seeds rotate through the sub-group selected for training.  Signals produced from each seed are averaged with equal weight across all seeds.  While QCML models can be updated online or through rolling / expanding retraining, for simplicity we keep the parameters static after initial training.  Periodic retraining or online updating of QCML parameters could likely lead to stronger results than those we present herein.

Our QCML models herein used a Hilbert space dimensionality of 12.  The training loop and loss function described in the algorithm in section \ref{section:qcml_review} was implemented in PyTorch, and models were trained using the Adam optimizer.  All the results and figures for this article were obtained on a laptop equipped with a 32-core 13th Gen Intel Core i9-13950HX CPU with 64GB of memory, supplemented by a NVIDIA RTX 5000 Ada Generation Laptop GPU.

Training a single QCML model on the laptop GPU takes approximately 15 seconds for the data and hyperparameters we selected.  After we have successfully trained our QCML models on the training data, QCML parameters remain static over the course of the test period (January 2014 through June 2024).  As explained in section \ref{section:distance_to_similarity}, we can then compute $S_{QCML}(\mathbf{x}_t)$ on each day the measure is needed.

Computing $S_{QCML}(\mathbf{x}_t)$ for the full test period, also leveraging the laptop GPU, takes approximately 50 seconds per QCML model.  This means, using only the laptop, all 50 seeds could be trained and final signals constructed in approximately 1 hour.  A cloud computing approach could run all seeds in parallel, providing a final signal in less than 2 minutes.  In this article, we use 50 seeds for our final results, but 20 seeds gives sufficient convergence for hyperparameter exploration, further reducing the computational burden.

\subsection{Momentum Spillover Signal Construction}
\label{section:signal_construction}

Once we have constructed a similarity matrix (denoted $S(\mathbf{x}_{t})$ for the $J \times J$ similarity matrix on date $t$ using a set of characteristics $\mathbf{x}_{t}$), we can create a simple momentum spillover signal.  The signal uses $S(\mathbf{x}_{t})$ to construct firm linkage weights, and the lagged returns of linked firms as additional inputs.  Let $\mathbf{r}_{t-l:t-1,i}$ represent the returns for firm $i$ from $t-l$ to $t-1$ for some prior number of days $l$, and let $w_{t,j,i}$ represent the signal input weight from firm $i$ for the signal to be computed for firm $j$ on $t$. We proceed as follows:

\begin{enumerate}
    \item Define $w_{t,j,i} = S(\mathbf{x}_{t})_{j,i} / \sum_i S(\mathbf{x}_{t})_{j,i}$
    
    \item The spillover signal using input return days $l$ on trading date $t$ for firm $j$ is: $f_{l,t,j} = \sum_i (w_{t,j,i} * \mathbf{r}_{t-l:t-1,i}$)
    
    \item We construct $f_{l,t,j}$ using both Euclidean and QCML similarity, for $l$ equal to 21 days, 63 days, 126 days, and 252 days.  We also construct composite (over $l$) signals which average each of these 4 input returns and use the resultant $f_{\bar{l},t,j}$ as input to signal production

    \item Each $f_{l,t,j}$ is individually demeaned by GICS Industry Code and cross-sectionally z-scored prior to signal evaluation

\end{enumerate}

\subsection{Signal Evaluation}
\label{section:signal_evaluation}

To test performance, we use covariance estimated daily from daily returns\footnote{Covariance is estimated using a technique proprietary to Duality Group, the details of which are not relevant for this article} to produce Markowitz optimal investment portfolios, where portfolio weight $w = V^{-1} R f$ given covariance $V$, spillover signal forecast $f$, and projection\footnote{If one wishes to solve for portfolio weights which maximize expected returns, with a penalty for expected portfolio variance, while maintaining zero exposure to a set of controls, then for weights $w$, spillover signal forecast $f$, asset variance $V$, risk aversion $\mu$, and controls $M$, one needs to solve for $w$ which minimizes $-w^\intercal f + 0.5 \mu w^\intercal V w$ such that $w^\intercal M = 0$. The solution is $w = V^{-1} R f$ where $R = I - M (M^\intercal V^{-1} M)^{-1} M^\intercal V^{-1}$.  Thus $R$ is a projection operator away from $M$, consistent with the desired investment process. $R f$ is also the residual from an inverse variance weighted regression of $f$ on $M$ \label{fnprojection}} operator $R$, to give risk optimal portfolio weights with no exposure to our controls.\footnote{Given a matrix of control features $M$, we have $f^\intercal M = 0$ and $w^\intercal M = 0$}  

Investment portfolios are smoothed over 21 days to proxy for the fact that realistic investment portfolios must trade into new forecasts gradually.  We compute daily returns to the smoothed portfolios, and analyze portfolio performance from January 2014 through June 2024.  We do not remove transaction costs as these forecasts are not intended to represent a stand-alone investment strategy.

\section{Results}

\subsection{Euclidean Similarity CVL Signals}

We first examine the properties and performance of the Euclidean similarity CVL signals.  In table \ref{table:euclidean_autocorrelation} we show the average daily half-lives of the signals\footnote{Define signal decay $d$ as the average cross-sectional correlation of $f_t$ with $f_{t-1}$.  We approximate half-life as $\ln{0.5} / \ln{d}$}.  Relative to the horizon of the input returns, half-lives are faster than would be expected from a simple average of returns, particularly for longer dated input returns.  This shows that linkages based on our accounting-data-focused inputs are fairly stable at a monthly horizon, but are an important source of variation over longer horizons.

\begin{figure}[ht!]
\caption{Cumulative returns to the Euclidean similarity CVL signals with varying input return horizons. Strategy portfolios are market neutral and have zero exposure to Analyst Connected Stock Momentum, Analyst Coverage, Beta, Momentum, Short Term Reversal, Size, Subindustry Momentum, Revenue to Price, and GICS Industry Groups, and thus have no linear contribution of those features to returns.  Returns have been scaled by full sample realized volatility of each time series.}
\centering
\includegraphics[width=0.55\textwidth]{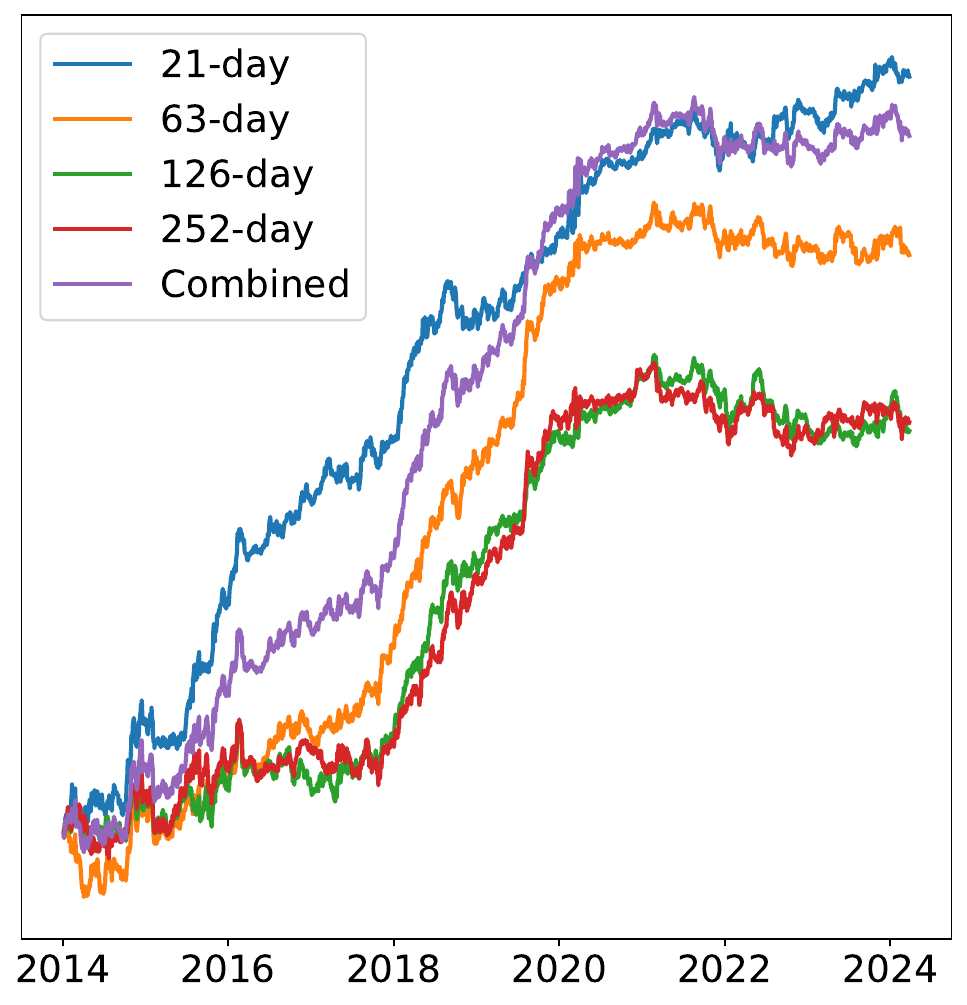}
\label{fig:euclidean_returns}
\end{figure}

In Table \ref{table:euclidean_sharpe} we show the Sharpe ratios of portfolios formed from each of the signals.  We plot the cumulative returns to the signals in Figure \ref{fig:euclidean_returns}.  Recall as explained in section \ref{section:signal_evaluation} that our market neutral investment portfolios are formed to have zero exposure to Analyst Connected Stock Momentum, Analyst Coverage, Beta, Momentum, Short Term Reversal, Size, Subindustry Momentum, Revenue to Price, and GICS Industry Groups, and thus have no linear contribution of those features to returns.

The Euclidean similarity signals all show positive performance on average, although performance significantly deteriorates in the final sub-period.  We also note that the signals with longer input horizon returns perform substantially worse than the signals with short input horizon returns.

\begin{table*}\centering
\caption{Average daily half-lives of the Euclidean similarity CVL signals with varying input return horizons.}
\begin{tabular}{@{}rccccc@{}}
\toprule
    \textbf{Period} & \textbf{21-day} & \textbf{63-day} & \textbf{126-day} & \textbf{252-day} & \textbf{Combined} \\ 
    \midrule
    Jan 2014 - Jun 2024 & 9.1 & 19.5 & 27.0 & 36.3 & 14.8 \\
    \arrayrulecolor{black!50} \midrule
    Jan 2014 - Jun 2017 & 9.2 & 19.7 & 26.9 & 35.6 & 14.9 \\
    Jul 2017 - Dec 2020 & 9.0 & 19.3 & 28.2 & 40.3 & 15.0 \\
    Jan 2021 - Jun 2024 & 9.0 & 19.4 & 25.9 & 33.5 & 14.5 \\
\arrayrulecolor{black} \bottomrule
\end{tabular}
\label{table:euclidean_autocorrelation}
\end{table*}

\begin{table*}\centering
\caption{Sharpe ratios of returns to the Euclidean similarity CVL signals with varying input return horizons.  Strategy portfolios are market neutral and have zero exposure to Analyst Connected Stock Momentum, Analyst Coverage, Beta, Momentum, Short Term Reversal, Size, Subindustry Momentum, Revenue to Price, and GICS Industry Groups, and thus have no linear contribution of those features to returns.}
\begin{tabular}{@{}rccccc@{}}
\toprule
    \textbf{Period} & \textbf{21-day} & \textbf{63-day} & \textbf{126-day} & \textbf{252-day} & \textbf{Combined} \\ 
    \midrule
    Jan 2014 - Jun 2024 &  1.35 &  1.03 &  0.71 &  0.73 &  1.24 \\
    \arrayrulecolor{black!50} \midrule
    Jan 2014 - Jun 2017 &  1.60 &  0.53 &  0.28 &  0.29 &  1.02 \\
    Jul 2017 - Dec 2020 &  1.63 &  2.50 &  2.00 &  2.00 &  2.41 \\
    Jan 2021 - Jun 2024 &  0.61 & -0.13 & -0.38 & -0.34 &  0.00 \\
\arrayrulecolor{black} \bottomrule
\end{tabular}
\label{table:euclidean_sharpe}
\end{table*}

\subsection{QCML Similarity CVL Signals}

We next examine the properties and performance of the QCML similarity CVL signals.  In table \ref{table:qcml_autocorrelation} we show the average daily half-lives of the signals.  The signal half-lives here follow a similar pattern to those obtained with the simpler Euclidean similarity in table \ref{table:euclidean_autocorrelation}, but with the important distinction that the half-lives of the signals here are universally longer, and to a greater extent the longer the horizon of the input returns.  In fact, the average half-life of the 252-day input return QCML similarity signal (90.9 days) is 2.5 times greater than the half-life of the 252-day input return Euclidean similarity signal (36.3 days).  This implies that rather than adding instability from a highly parameterized learning approach, QCML has likely managed to focus on the more resilient relationships which can be inferred from the input data.

\begin{figure}[ht!]
\caption{Cumulative returns to the QCML similarity CVL signals with varying input return horizons. Strategy portfolios are market neutral and have zero exposure to Analyst Connected Stock Momentum, Analyst Coverage, Beta, Momentum, Short Term Reversal, Size, Subindustry Momentum, Revenue to Price, and GICS Industry Groups, and thus have no linear contribution of those features to returns.  Returns have been scaled by full sample realized volatility of each time series.}
\centering
\includegraphics[width=0.55\textwidth]{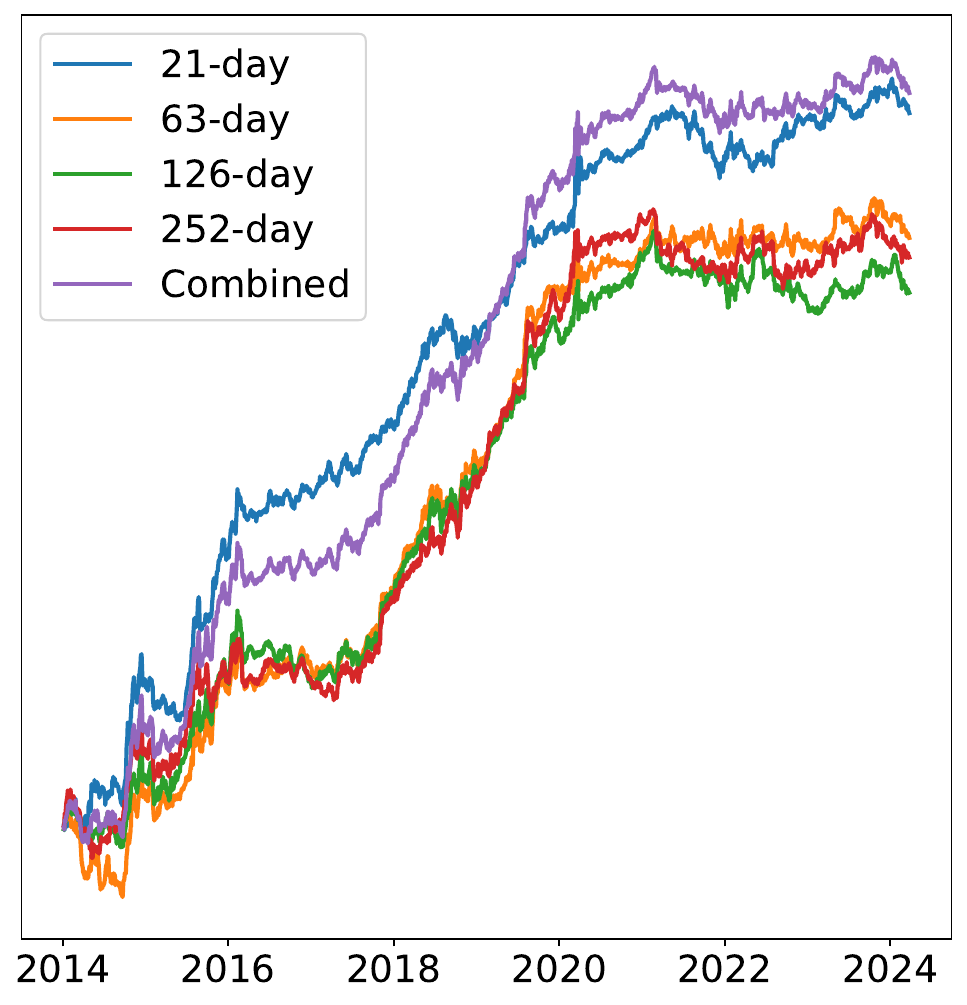}
\label{fig:qcml_returns}
\end{figure}

In Table \ref{table:qcml_sharpe} we show the Sharpe ratios of portfolios formed from each of the signals.  We plot the cumulative returns to the signals in Figure \ref{fig:qcml_returns}.  Recall as explained in section \ref{section:signal_evaluation} that our market neutral investment portfolios (as for the Euclidean similarity investment portfolios) are formed to have zero exposure to Analyst Connected Stock Momentum, Analyst Coverage, Beta, Momentum, Short Term Reversal, Size, Subindustry Momentum, Revenue to Price, and GICS Industry Groups, and thus have no linear contribution of those features to returns.

Performance is similar to that obtained with the simpler Euclidean similarity in table \ref{table:euclidean_sharpe} (the averages of the daily cross-sectional correlations of the underlying signals range from 0.74 to 0.78).  However, the signals derived from QCML similarity all show out-performance, and to a greater extent as the horizon of the input returns increases.  While the 21-day input return QCML only slightly outperforms, the 252-day input return QCML signal shows a Sharpe of 1.10 compared to 0.73 for the Euclidean signal (a greater than 50\% improvement in Sharpe), despite the fact the QCML signal has a half-life (90.9 days) 2.5 times longer than the Euclidean signal (36.3 days).  The QCML combined input return signal gives a Sharpe of 1.42, an improvement over the Euclidean combined input return signal, which has a Sharpe of 1.24, while the QCML signal has a half-life (23.5 days) nearly 1.6 times longer than the Euclidean signal (14.8 days).  

Additionally, the returns to the majority of the QCML similarity signals show somewhat better sub-period consistency.  These findings again imply that QCML has likely managed to learn more resilient cross-firm relationships than those computed using the simpler Euclidean similarity. Note that these results have also been obtained by training QCML a single time, on data prior to the test period.  Periodic retraining or online updating of QCML parameters could likely lead to further improved results.

\begin{table*}\centering
\caption{Average daily half-lives of the QCML similarity CVL signals with varying input return horizons.}
\begin{tabular}{@{}rccccc@{}}
\toprule
    \textbf{Period} & \textbf{21-day} & \textbf{63-day} & \textbf{126-day} & \textbf{252-day} & \textbf{Combined} \\ 
    \midrule
    Jan 2014 - Jun 2024 & 11.5 & 32.6 & 53.5 &  90.9 & 23.5 \\
    \arrayrulecolor{black!50} \midrule
    Jan 2014 - Jun 2017 & 12.5 & 34.9 & 54.6 &  90.2 & 25.0 \\
    Jul 2017 - Dec 2020 & 11.5 & 33.3 & 61.5 & 120.5 & 25.4 \\
    Jan 2021 - Jun 2024 & 10.6 & 29.7 & 46.0 &  72.2 & 20.5 \\
\arrayrulecolor{black} \bottomrule
\end{tabular}
\label{table:qcml_autocorrelation}
\end{table*}

\begin{table*}\centering
\caption{Sharpe ratios of returns to the QCML similarity CVL signals with varying input return horizons.  Strategy portfolios are market neutral and have zero exposure to Analyst Connected Stock Momentum, Analyst Coverage, Beta, Momentum, Short Term Reversal, Size, Subindustry Momentum, Revenue to Price, and GICS Industry Groups, and thus have no linear contribution of those features to returns.}
\begin{tabular}{@{}rccccc@{}}
\toprule
    \textbf{Period} & \textbf{21-day} & \textbf{63-day} & \textbf{126-day} & \textbf{252-day} & \textbf{Combined} \\ 
    \midrule
    Jan 2014 - Jun 2024 &  1.38 &  1.14 &  1.04 &  1.10 &  1.42 \\
    \arrayrulecolor{black!50} \midrule
    Jan 2014 - Jun 2017 &  1.84 &  0.89 &  0.88 &  0.78 &  1.47 \\
    Jul 2017 - Dec 2020 &  1.68 &  2.14 &  2.14 &  2.48 &  2.31 \\
    Jan 2021 - Jun 2024 &  0.28 &  0.14 & -0.25 & -0.28 &  0.08 \\
\arrayrulecolor{black} \bottomrule
\end{tabular}
\label{table:qcml_sharpe}
\end{table*}

\section{Conclusion}

We introduced a novel form of firm linkages, which we call Characteristic Vector Linkages (CVLs).  We then demonstrated how both simple Euclidean similarity CVLs and Quantum Cognition Machine Learning (QCML) similarity CVLs can be leveraged to capture momentum spillover effects in equity markets.

We find that CVLs constructed from vectors of firm characteristics provide a meaningful basis for identifying economically linked firms, evidenced by the positive Sharpe ratios achieved across the various input return horizons and similarity types, displayed in tables \ref{table:euclidean_sharpe} and \ref{table:qcml_sharpe}.

The Euclidean similarity approach, despite its simplicity, generates risk controlled portfolio returns with Sharpe ratios ranging from 0.73 to 1.37 depending on the input horizon, with shorter-horizon signals generally outperforming longer-horizon variants.

As discussed in detail in \cite{Candelori_et_al_2025}, QCML learns non-commutative quantum models for the data manifold. These quantum models have the ability to abstract out the fundamental features of the geometry of the data manifold, in a manner that is robust to noise.  QCML therefore offers an elegant framework for representing complex, non-linear patterns in data, and subsequently making more accurate measurements of distance between data points.  In this paper, we take advantage of QCML's abilities to ultimately learn relationships between firms that may not be captured by traditional distance metrics. By mapping firm characteristics into quantum states and leveraging quantum fidelity to derive similarity measures, QCML appears to uncover more accurate connections that enhance the predictive power of momentum spillover strategies.

Our supervised QCML similarity approach demonstrates meaningful improvements over the simple Euclidean method, particularly for longer-horizon signals. While the performance differences are modest for short-term signals, with 252-day input returns the QCML similarity measure achieves a Sharpe ratio of 1.12 compared to 0.76 for the corresponding Euclidean measure, while having a half-life (90.9 days) which is 2.5 times longer than the Euclidean signal's half-life (36.3 days). The combined QCML signal delivers a Sharpe ratio of 1.43, an  improvement over the Euclidean combined signal's 1.26 Sharpe ratio, while having a half-life (23.5 days) which is nearly 1.6 times longer than the Euclidean signal's half-life (14.8 days).

The QCML-based signals also exhibit somewhat superior consistency across sub-periods. While both approaches perform poorly during the January 2021 through June 2024 period, the QCML signals do modestly outperform the Euclidean signals. The QCML signals also derive relatively less of their overall performance from the strongest performing second sub-period (July 2017 through December 2020).  This improved stability, as well as the considerably longer average half-lives observed for QCML signals, suggest that the supervised learning approach successfully identifies more robust and persistent firm relationships.  Additionally, these results have been obtained by training QCML a single time, on data prior to the test period.  Periodic retraining or online updating of QCML parameters could likely lead to stronger outperformance during the most recent sub-period.  There is no such possible avenue to improve the performance of the Euclidean similarity signals.

Our work in this article could be extended in two obvious ways.  First, as noted in section \ref{section:setup}, our QCML models are only trained a single time, on data available prior to the test period.  A more dynamic approach allowing for periodic retraining or online updating of parameters may give improved estimates of similarity.  Second, as mentioned in section \ref{section:data}, we did not take a view regarding which sort of firm characteristics are most likely to be meaningful for capturing CVLs.  Future research could hypothesize economic mechanisms driving the ability of CVLs to capture firm linkages, and thereby refine the categories of input factors used or the target selected.  The simplicity of the CVL approach makes testing such data variations trivial.

In conclusion, this work introduces Characteristic Vector Linkages as a promising new approach for capturing firm relationships and demonstrates the potential for Quantum Cognition Machine Learning to more accurately measure distance between data points and thus capture similarity relationships. As markets continue to evolve and traditional strategies face increasing competition, the ability to identify and exploit subtle firm linkages through advanced similarity learning may become increasingly valuable for quantitative investment strategies.

\bibliographystyle{plain}
\bibliography{refs}

\end{document}